\documentclass[fleqn,11pt]{wlscirep}
\usepackage[utf8]{inputenc}
\usepackage[T1]{fontenc}
\usepackage{natbib}
\hypersetup{
colorlinks = true,
linkcolor = [rgb]{0.70,0.13,0.13},
citecolor = [rgb]{0.13,0.55,0.13},
urlcolor = [rgb]{0.25, 0.41, 0.88}}

\usepackage{hyperref}       
\usepackage{url}            
\usepackage{booktabs}       
\usepackage{amsfonts}       
\usepackage{nicefrac}       
\usepackage{microtype}      
\usepackage{xcolor}         
\usepackage{bbm}
\usepackage{soul}
\usepackage{graphicx}
\usepackage{bm}
\usepackage{amsmath}
\DeclareMathOperator*{\argmax}{argmax}

\usepackage{subcaption}
\usepackage[export]{adjustbox}

\title{Active Learning for Discovering Complex Phase Diagrams with Gaussian Processes}
\author[1,$\dagger$]{Max Zhu}
\author[2,$\dagger$]{Jian Yao}
\author[3]{Marcus Mynatt}
\author[3]{Hubert Pugzlys}
\author[3]{Shuyi Li}
\author[1]{Sergio Bacallado} 
\author[1,*]{Qingyuan Zhao}
\author[3,*]{Chunjing Jia}
\affil[1]{Statistical Laboratory, University of Cambridge, Cambridge, CB3 0WA, United Kingdom}
\affil[2]{Department of Physics, Southern University of Science and Technology, Shenzhen, 518055, China}
\affil[3]{Department of Physics, University of Florida, 2001 Museum Road, Gainesville FL 32611, United States}

\affil[$\dagger$]{Equal Contribution}
\affil[*]{ qyzhao@statslab.cam.ac.uk; chunjing@ufl.phys.edu}

\begin{abstract}
We introduce a Bayesian active learning algorithm that efficiently elucidates phase diagrams. Using a novel acquisition function that assesses both the impact and likelihood of the next observation, the algorithm iteratively determines the most informative next experiment to conduct and rapidly discerns the phase diagrams with multiple phases. Comparative studies against existing methods highlight the superior efficiency of our approach. We demonstrate the algorithm’s practical application through the successful identification of the entire phase diagram of a spin Hamiltonian with antisymmetric interaction on Honeycomb lattice, using significantly fewer sample points than traditional grid search methods and a previous method based on support vector machines. Our algorithm identifies the phase diagram consisting of skyrmion, spiral and polarized phases with error less than $5\%$ using only $8\%$ of the total possible sample points, in both two-dimensional and three-dimensional phase spaces. Additionally, our method proves highly efficient in constructing three-dimensional phase diagrams, significantly reducing computational and experimental costs. Our methodological contributions extend to higher-dimensional phase diagrams with multiple phases, emphasizing the algorithm's effectiveness and versatility in handling complex, multi-phase systems in various dimensions.

\end{abstract}

\begin{document}

\flushbottom
\maketitle

\thispagestyle{empty}

\section{Introduction and Background}
The phase diagram represents an indispensable map for understanding the different states of matter and the phase transitions between them under varying conditions, such as temperature, pressure, chemical concentration etc. This conceptual tool is not only fundamental in thermodynamics but also plays a pivotal role in material physics for the  understanding of the interplay of different microscopic degrees of freedom. On the practical side, it aids the design of functional materials and the exploration of emergent phenomena for next-generation energy materials, guiding the synthesis and characterization of materials such as alloys \cite{alloyphase}, perovskites, and strongly correlated electronic systems \cite{DagottoReview}. For instance, in the realm of unconventional superconductivity, phase diagrams serve a critical role in identifying the conditions (temperature, doping concentration, pressure, etc) under which materials exhibit superconductivity and display the competing phases such as antiferromagnetism, fermi liquid phase, etc. \cite{superconductor}. In the design of perovskite solar cells, which represent a promising photovoltaic technology
, the ability to design new halide perovskite systems could rely on stabilizing the desired functional phases within specific strain or pressure ranges. \cite{PerovskiteStrain, PerovskitePressure} Likewise, in the realm of magnetic materials, comprehending the phase diagram across various microscopic parameters such as spin exchange and Kitaev interactions, as well as external magnetic field, is essential for exploring novel magnetic states of matter, including quantum spin liquids, non-collinear magnetism, and magnetic skyrmions, etc. Thus, the phase diagram serves as an invaluable reference for scientists and engineers striving to manipulate the physical states of matter toward innovative ends, and efficiently determining the phase diagram for complex systems across different materials systems is crucial.

Physicists and materials scientists have invested significant effort in developing state-of-the-art experimental techniques and numerical simulations to obtain the phase for each parameter set. To obtain the entire phase diagram, the traditional approach, whether experimental or theoretical, typically entails a time-consuming grid search over the entire phase space. However, this process can become highly inefficient as the dimension of the parameter space and the number of phases increase, and it can become prohibitively expensive or time-consuming when sophisticated experiments or simulations are needed to evaluate the phase. To our knowledge, despite much methodological progress about experimental design in statistics and machine learning, little attention has been given to the problem of discovering complex phase diagrams until recently \cite{mlforpdGuillaume2022, mlforpdLund2022, mlforpdLink2023}. Here we present an active learning framework to increase sampling efficiency.

Active learning \cite{activeSurvey} is an iterative process in which an algorithm uses existing data to determine how to sample new observations. Originally proposed for complex optimization problems~\cite{EI}, active learning has found applications in a broad range of problems including optimizing material design \cite{activeMaterials}. Our work is motivated by the active learning method in \citet{efficientPhase} that uses a Gaussian Process (GP) regression \cite{gpForML} to prioritize sampling new points near the predicted phase boundary and of high uncertainty. However, the acquisition function in \citet{efficientPhase} is based on a simple heuristic and does not fully use the GP posterior, and their method is limited to modeling diagrams with just two phases. Another recent work by \citet{multiPhase} investigated phase diagrams with more than two phases using a classifier to predict phases. They propose two methods, one based on Support Vector Machines that does not quantify the uncertainty of boundaries, and a second method that quantifies uncertainties but requires direct observation of the phase transition point at the phase boundary, rather than relying on samples taken from inside each phase. Our proposed method can quantify uncertainties without requiring direct observations of phase boundaries. 

We propose a general Bayesian active learning framework for efficiently discovering phase diagrams. Our method is particularly adept at addressing phase diagram challenges encountered in physics and materials science, where each phase is contiguous in parameter space and the occurrence of isolated "islands" of phase within another phase is improbable. Compared to limited existing approaches, our method has the following features: (1) the phase diagram can contain any number of phases; (2) the acquisition function is purposefully designed using the GP posterior to estimate the expected change in phase diagram after sampling a new point; (3) we do not require direct measurements the phase boundary and allow noisy measurements of the phase at the sampled parameter values. As a result, our algorithm converges to the true phase diagram considerably faster than existing methods in simulations.

We apply our algorithm to investigate the phase diagram of a spin system with antisymmetric interaction on a monolayer honeycomb lattice under the influence of a magnetic field. Skyrmions, as an exotic state of matter with topological spin configuration, hold significant promise for next-generation technological applications in areas such as data storage, spintronics, sensor technology, and quantum computation, etc.\cite{review2, review3}  
Our algorithm efficiently identifies the skyrmion phase, which is sandwiched between a spiral phase and a polarized phase, using only a small sample in the Hamiltonian's high-dimensional parameter space.
Specifically, our algorithm identifies the entire phase diagram, including skyrmion, spiral, and polarized phases, with an error of less than $5\%$, using only $8\%$ of the total possible sample points. The results are robust in both two-dimensional and three-dimensional phase spaces. As demonstrated by the effective sampling for obtaining the entire phase diagram, we show that our active learning algorithm has the great potential to revolutionize the time-consuming phase diagram inquiry if it gets embedded with the automated sampling in phase space for sophisticated experiments or simulations. This integration could significantly accelerate the discovery of emergent phases and the design of functional materials, particularly those with high-dimensional tuning parameters or where each experiment or simulation is time-intensive.

\section{Methodology}

Our objective is to model a phase diagram with $n$ phases (e.g. state of matter) determined by $k$ parameters
$\mathbf{x} \in \mathbb{R}^k $ (e.g. temperature or interaction strength), where the true phase $y(\mathbf{x}) \in \{1,...,n\}$ varies with $\mathbf{x}$. 
We assume the user can conduct experiments (or simulations) at a point $\mathbf{x}$ to give a possibly noisy measurement $\tilde{\mathbf{y}} = (p_1, .., p_n) \in \mathbb{R}^n$, where $p_m=p(y=m \mid \mathbf{x} = \mathbf{x}_{t+1})$ is the probability of phase $m$ being the true phase at point $\mathbf{x}$. 
Our active learning algorithm takes existing observations $\mathcal{T} = \{\mathbf{x}_i, \tilde{\mathbf{y}}_i\}^{t}_{i=1}$ at step $t$ to determine the next point $\mathbf{x}_{t+1}$ to sample, and the user conducts an experiment at $\mathbf{x}_{t+1}$ to give observation $\tilde{\mathbf{y}}_{t+1}$ that is added to the dataset. This process is repeated until a phase diagram with satisfactory precision is obtained. A good active learning algorithm would suggest new points that allow the user to discover the phase diagram with as few measurements as possible.

\paragraph{Observations}
For noiseless observations, the phase is exactly measured so the measurement posterior consists of a one-hot vector $\tilde{\mathbf{y}}(\mathbf{x}) = (0, ..., p_m=1, ..., 0 )$ if the measured phase is $y=m$. 
For noisy observations, $\tilde{\mathbf{y}}$ is the estimated probability of the true phase, including experimental error. However, real-world experiments often do not directly return a full distribution. In this case, $\tilde{\mathbf{y}}$ is the measurement posterior distribution derived from the experimental outcome and statistical error. For example, each experiment may return a single phase that is incorrect with probability $\epsilon$ that can be estimated from prior knowledge, e.g. through analysis of experimental uncertainty or knowledge about experimental noise levels. If phase $m$ is deemed most probable in this case, the vector $\tilde{\mathbf{y}}$ could consist of $p_m = 1-\epsilon$ and $p_j = \epsilon/(n-1), j \neq m$ for all other phases. Experimental error can vary with experimental parameters, $\epsilon = \epsilon(\mathbf{x})$ to allow for increased experimental errors near the phase boundary. This flexible setup allows the user to incorporate more detailed knowledge of measurement errors, for example, when multiple experiments are conducted at a single parameter value or if different phases have different measurement errors.

\paragraph{Gaussian Process Regression}
To predict the true phase at unmeasured parameter values and quantify its uncertainty, we use a Gaussian Processes (GP) regression model. We give a brief review of GP regression here and a more detailed introduction can be found in \cite{gpForML}. 

A Gaussian process (or a Gaussian random field, when the dimension $k > 1$) is a collection of random variables $\{a(\mathbf{x}):\mathbf{x} \in \mathcal{X}\}$ such that for any finite set of indices $\mathbf{x}_1,\dots,\mathbf{x}_m \in \mathcal{X}$, the marginal distribution of $(a(\mathbf{x}_1), \dots, a(\mathbf{x}_m))$ is multivariate normal. In GP regression, we observe a set of observed responses $\mathbf{a} = (a(\bm x_1),..,a(\bm x_n))$ at points $\mathbf{X} = (\mathbf{x}_1, \dots, \mathbf{x}_n)$ and we would like to predict the responses $\mathbf{a}_*=(a(\mathbf{x}_{n+1}), \dots, a(\mathbf{x}_{n+n_*}))$ at new points $\mathbf{X}_* = (\mathbf{x}_{n+1}, \dots, \mathbf{x}_{n+n_*})$. The GP prior assumes that the responses follow a Gaussian process with covariance kernel function $K$, that is, the covariance between two points $a(\mathbf{x}_i)$ and $a(\mathbf{x}_j)$ is given by $\Sigma_{ij} = K(\mathbf{x}_i, \mathbf{x}_j)$. The prior joint distribution is (assuming zero prior mean): 
\begin{equation}
\begin{bmatrix}
\mathbf{a} \\
\mathbf{a_*}
\end{bmatrix}
\sim \mathcal{N}\left(
\begin{bmatrix}
\mathbf{0} \\
\mathbf{0}
\end{bmatrix}, 
\begin{bmatrix}
{\Sigma} & {\Sigma}_* \\
{\Sigma}_*^T & {\Sigma}_{**}
\end{bmatrix}
\right),
\end{equation}
where ${\Sigma} = K(\mathbf{X}, \mathbf{X})$ is the matrix of the covariances evaluated between all pairs of known points $\mathbf{X}$ and ${\Sigma}_* = K(\mathbf{X}_*, \mathbf{X})$ and ${\Sigma}_{**} = K(\mathbf{X}_*, \mathbf{X}_*)$ are defined similarly. When $\mathbf{a}$ contains noisy measurements of the "true" response, a diagonal matrix can be added: ${\Sigma} = K(\mathbf{X}, \mathbf{X}) + \sigma^2 I$ where $\sigma^2$ is the noise level. The posterior distribution of $\mathbf{a}_*$ is then given by 
\begin{equation} \label{eq:GPPosterior}
\mathbf{a}_* \mid \mathbf{a} \sim \mathcal{N}(\Sigma_*^T \Sigma^{-1} \bm{a}, \Sigma_{**} - \Sigma_*^T \Sigma^{-1} \Sigma_*).
\end{equation}
Since this posterior is normal, the uncertainty of the prediction is quantified by the covariance matrix. 

In the application below 
we choose the Matern 1/2 kernel \citep{MaternKernel} as the covariance kernel $K$. This is a stationary kernel, so the covariance between two points depends only on the distance between them. The kernel has two parameters---length scale and variance---that are fitted using maximum marginal likelihood estimation (along with $\sigma^2$ if the response is noisy). To avoid cluttering, we will drop the stars in the posterior distributions below.

\paragraph{Our Model} We next describe how we fit the GP regression model to phase measurements. We model the probability distribution of the measured phase diagram by a softmax function of $a_m(\mathbf{x}), m \in \{1...n\}$, where $a_m(\mathbf{x})$ is the logit for phase $m$ at point $\mathbf{x}$. The logits are combined to give the predicted observation probability for each phase $m$ as 
\begin{equation} \label{eq:Softmax}
p(y = m \mid \mathbf{x}, a_1,\dots,a_n) = \frac{\text{exp}({a_m})}{\sum_{j=1}^{n}\text{exp}({a_j})},
\end{equation}
so a higher logit value (relative to other phases) gives a higher probability. The logits $a_1(\mathbf{x}),\dots,a_m(\mathbf{x})$ are assumed to be independent GPs a priori. This allows us to fit the GP regression for each phase separately as described below. 

Since our dataset $\mathcal{T}$ consists of a series of probabilities for every phase, we transform the probabilities into logits using the inverse of Equation \ref{eq:Softmax}: 
\begin{equation} \label{eq:SoftmaxSolve}
    a_m = \text{log}(p_m) - \frac{1}{n} \sum_{j=1}^n \text{log}(p_j).
\end{equation}
The logits can be scaled by an arbitrary additive constant, so we choose to normalize them to sum up to 0. This is consistent with the choice of zero prior mean for the GPs. However, a practical issue with equation \eqref{eq:SoftmaxSolve} is that noiseless phase measurements with $p_m \in \{0, 1\}$ result in infinite logits. To address this, we clip the probabilities between $0.01/(n-1)$ and $0.99$. 

Next, we fit GP regression models on the calculated logits to give a posterior distribution of the logits 
using equation \eqref{eq:GPPosterior}. The posterior distribution of the phase at point $\mathbf{x}$ is calculated as
\begin{equation} \label{eq:IntProb}
p(y = m \mid \mathbf{x}, \mathcal{T}) = \int_{\mathbb{R}^n} p(y = m \mid \mathbf{x}, a_1,...,a_n) \prod_j p(a_j \mid \mathbf{x}, \mathcal{T}) d\mathbf{a},
\end{equation}
so the the uncertainty about the logits is properly aggregated. If all logits have a high variance (i.e., low certainty), all the probabilities will be around $1/n$. The integral in equation \ref{eq:IntProb} is numerically approximated using Gauss-Hermite quadrature \citep{GaussHermite} and 
can be interpreted as a probabilistic prediction of the experimental outcome at point $\mathbf{x}$. The predicted phase is taken as the phase with the highest predicted probability, 
\begin{equation} \label{eq:PredictPhase}
    P_\mathcal{T}(\mathbf{x}) = \argmax_m p(y=m \mid \mathbf{x}, \mathcal{T}).
\end{equation}
Evaluating this over a grid generates a predicted phase diagram, $P_\mathcal{T}$. The uncertainty at a given point can be characterized by the probability that the true phase is anything other than the most likely phase. This predicted phase diagram $P_\mathcal{T}$ is taken as the model predictions, while the predicted uncertantity is $1 - \argmax_m p(y=m \mid \mathbf{x}, \mathcal{T})$, the probability of measuring any other phase. 

\begin{figure}
    \centering
    \includegraphics[width=0.9\textwidth]{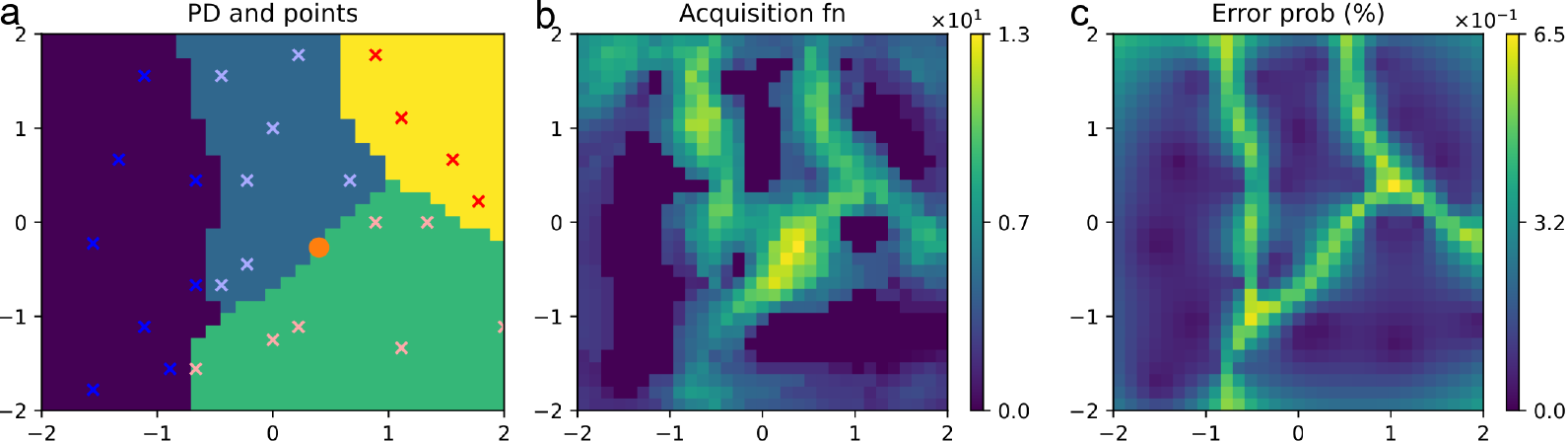}
    \caption{Visualization of our model internal state on a phase diagram with four phases. (\textbf{a}) Predicted most likely phase (shaded) with previous observations (crosses), colored by the phase observed. The recommended point to sample is shown as an orange dot. (\textbf{b}) Acquisition function rating points to sample. (\textbf{c}) Predicted error (\%) of observing different phases from predicted. }
    \label{fig:ourPredictions}
\end{figure}

\paragraph{Acquisition function} 
In active learning, the next point to sample $\mathbf{x}_{t+1}$ is determined in a greedy fashion by maximizing the acquisition function $A(\mathbf{x})$ that measures how much additional information a new measurement at $\mathbf{x}$ can provide: $\mathbf{x}_{t+1} = \argmax_\mathbf{x} A(\mathbf{x})$. 
A good choice of acquisition function leads to a more efficient experimental design.

We propose to use expected change in area/volume of the phase diagram as the acquisition function. Consider the following distance function between two phase diagrams $P_1$ and $P_2$: (let $V$ be the total volume of the phase space)
\begin{equation}
    D(P_1,P_2) = \frac{1}{V} \int \mathbbm{1}(P_1( \textbf{x})\neq P_2( \textbf{x})) d\textbf{x}
\approx \frac{1}{|\mathcal{G}|} \sum_{\textbf{x} \in \mathcal{G}} \mathbbm{1}(P_1( \textbf{x})\neq P_2(\textbf{x})).
\end{equation}
where $\mathcal{G}$ is a pre-chosen grid of points used for numerical approximation. 
Our acquisition function is defined as 
\begin{equation} \label{eq:AcqFn}
    A(\mathbf{x}) = \mathbb{E}_{\hat{\mathbf{y}}}[D(P_\mathcal{T}, P_{\mathcal{T} \cup \{\mathbf{x}, \hat{\mathbf{y}}\}})],
\end{equation} 
the expected distance between the current most probable phase diagram $P_\mathcal{T}$ and the most probable phase diagram after a "fantasy experiment" is performed at point $\mathbf{x}$. The expectation in Equation \eqref{eq:AcqFn} is taken over the posterior distribution of the "fantasy measurement" $\hat{\mathbf{y}}$ at $\mathbf{x}$ given by the GP regression model. "Fantasy experiments" are hypothetical observations that are distributed according to our current model (Equation \eqref{eq:IntProb}) that simulate the outcome of real experiments, 
 
\begin{equation} \label{eq:FantasyProbs}
    \hat{\mathbf{y}} = (\hat{\epsilon}/n, ..., p_m=1-\hat{\epsilon}, ..., \hat{\epsilon}/n), \quad m \sim p(y|\mathbf{x}, \mathcal{T}),
\end{equation}
where $\hat{\epsilon}$ is a user chosen hyperparameter representing a rough estimate of experimental noise and the main measured phase $j$ is distributed according to our model. Ideally, $\hat{\epsilon}$ would equal the real experimental noise, but this is unknown so we set it to a constant using prior knowledge about experimental noise. For the noiseless case, we choose $\hat{\epsilon} = 0.01$ because setting it too low may result in numerical instability. Fantasy observations mirror the setup described in the \textit{Observations} section. 

To compute $A(\mathbf{x})$, we fit our model on our current observations $\mathcal{T}$ and generate the predicted phase diagram ${P}_\mathcal{T}$ using Equation \ref{eq:PredictPhase}. Then, for each phase $m$, we fit a new model on the current dataset with an extra fantasy observation, $\{\mathbf{x}, \mathbf{\hat{y}} = \mathbf{\hat{y}}_m\}$, where $\mathbf{\hat{y}}_m$ is the fantasy observation probabilities for phase $m$ (Equation \ref{eq:FantasyProbs}), and generate predictions giving $P_{\mathcal{T} \cup \{\mathbf{x}, \mathbf{\hat{y}}_m\}}$. Finally, $A(\mathbf{x})$ is computed by taking the mean distance between the current phase diagram and each possible fantasy phase diagram, weighted by the probability that each fantasy diagram occurs. 

This acquisition function finds points that will maximally change the phase diagram, by combining the impact of an observation on the entire phase diagram with the probability of that observation and balances the exploration of unknown regions with the refinement of known boundaries. Compared to other commonly used acquisition functions, such as KL divergence or Expected Error Reduction \citep{ActiveEER}, our acquisition function prioritizes discovering the phase boundary while placing less emphasis on points far away from the phase boundary. The method in \citet{efficientPhase} differs from ours as it finds the points that are locally maximally uncertain, instead of considering the impact of an observation on a region surrounding the point. This process is visualized in Figure \ref{fig:ourPredictions}. Our acquisition function is shown in Figure \ref{fig:ourPredictions}(b), while Dai and Glotzer would pick a point that maximizes the uncertainty in Figure \ref{fig:ourPredictions}(c).

\begin{figure}
    \centering
    \includegraphics[width=1.00\linewidth, valign=t]{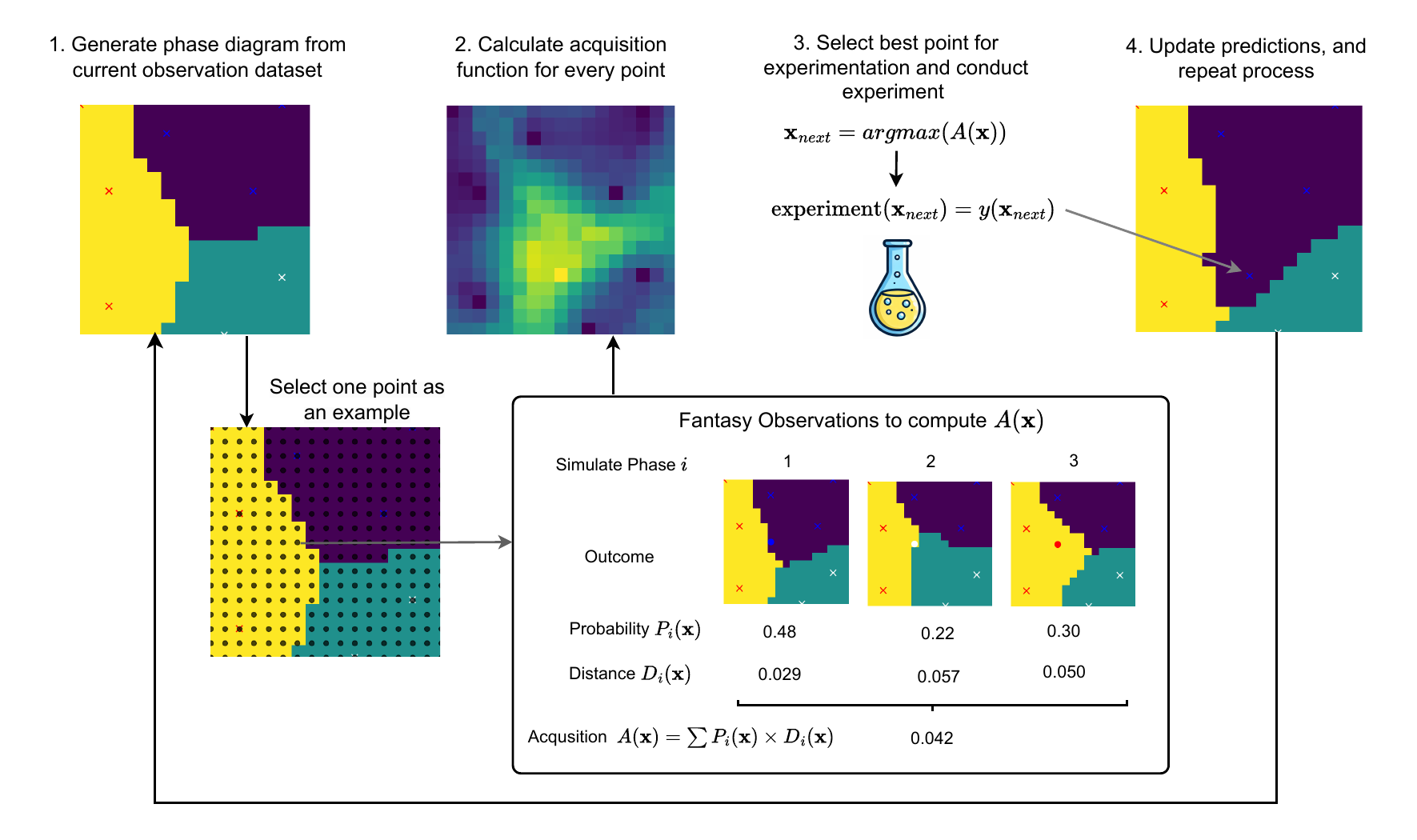}
    \caption{A schematic flowchart of the algorithm for our proposed active learning approach to phase diagram problems.}
    \label{fig:schematic}
\end{figure}

\paragraph{Implementation Details} 
We summarize the processes of our proposed active learning approach for phase diagram identification in algorithmic flowchart as shown in Figure \ref{fig:schematic}. We implement our method using PyTorch \citep{PyTorch}. Appendix \ref{Appendix:details} describes implementation details and several numerical optimizations to speed up the algorithm. With these optimizations, each step of active learning requires about 25 seconds on an 8-core processor for a three-dimensional phase diagram.

\section{Synthetic Experiments} \label{sect:Synthetic}
\begin{figure}
    \centering
    \includegraphics[width=\textwidth]{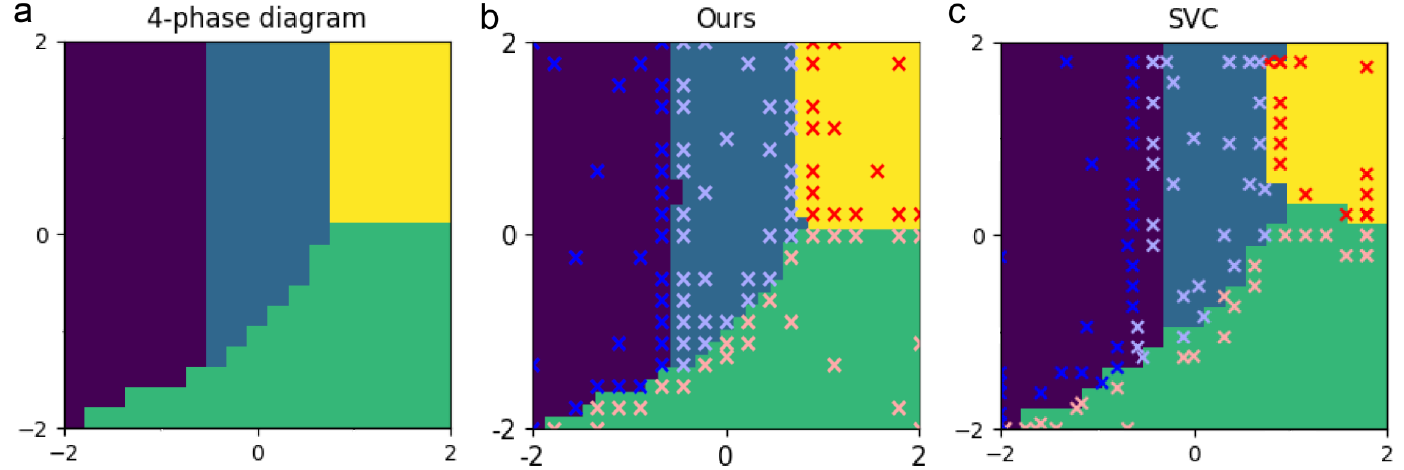}
    \caption{(\textbf{a}) A phase diagram of four phases used for testing, with phases shown with shaded color. Predicted phase diagram after 100 observations from our model (\textbf{b}) and SVC (\textbf{c}), with predicted phase shaded and observations as crosses colored by observed phase. Note the SVC repeats a large number of observations and does not attempt to search the green region for undiscovered phases. }
    \label{fig:quad_pd}
\end{figure}
Our model is evaluated against \citet{efficientPhase} (Gaussian process regression or GPR) 

and a modified version of Approach 1 in \citet{multiPhase} (support vector machine classification or SVC)

as baselines on two synthetic examples. Like our method, GPR is a Bayesian Gaussian Process method but is only suitable for binary classification and cannot directly handle experimental uncertainty. The SVC method uses a support vector machine to predict classification boundaries and places observations farthest from other observations on the boundary. The SVC cannot account for uncertainty, quantify prediction uncertainty or discover unknown isolated phases since the model will not search regions not on the boundary. Further details of these baselines are given in Appendix \ref{Appendix:baselines}. These experiments are restricted to $k=2$ for easier analysis and visualization. The first experiment is on a noisy phase diagram, while the second has several phases. 

\begin{figure}
    \centering
    \includegraphics[width=0.75\linewidth, valign=t]{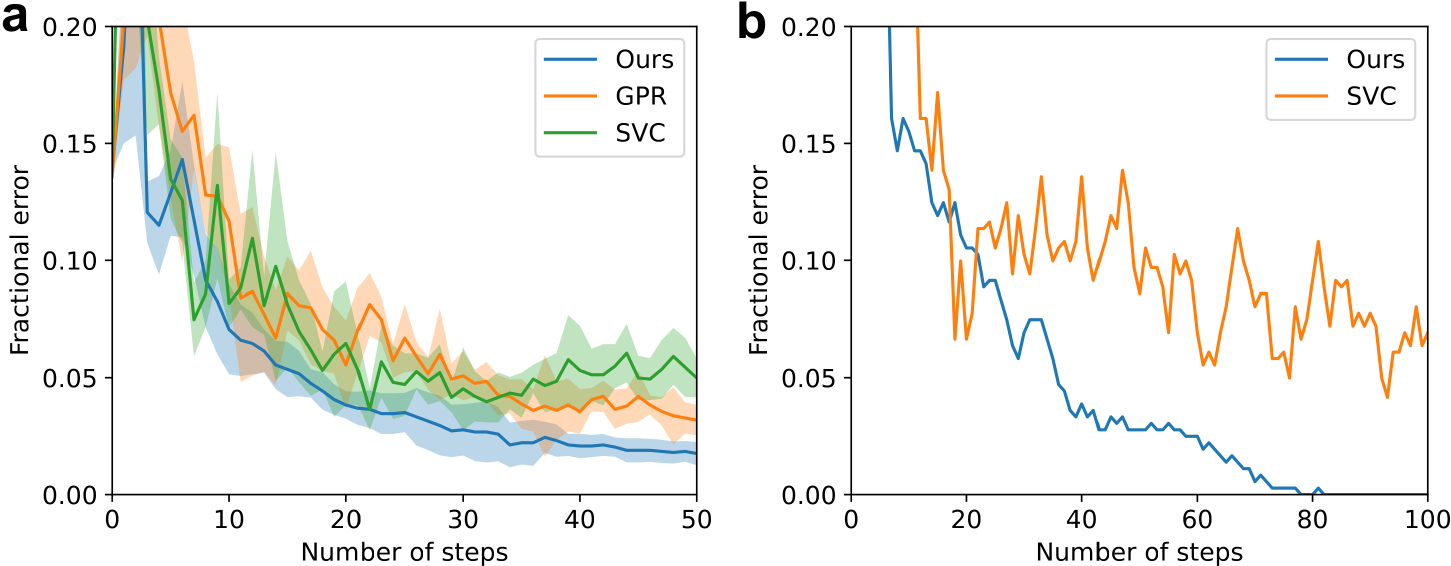}
    \caption{Fractional error between predicted and true phase diagrams versus number of samples taken for phase diagrams of (\textbf{a}) two phases and (\textbf{b}) four phases. Shaded area in (\textbf{a}) shows standard deviation over 6 runs with different random noise, while (\textbf{b}) is noiseless.}
    \label{fig:acc_plot}
\end{figure}

\textbf{Two phase problem}: Firstly, we evaluated the models on a noisy toy phase diagram of two phases, $P(x_1,x_2) = \mathbbm{1}(x_2 > \sin(0.5 \pi x_1))$  where $x_1, x_2 \in [-2, 2]$ represent experimental parameters. Uniform noise, $\epsilon\sim[-0.2, 0.2] $ was added to $x_1, x_2$ every observation to simulate experimental errors. Models were initialized with 2 observations. We assume measurement error is known for this experiment, so we input the error probability into our model, while the baselines cannot use the measurement error. Figure \ref{fig:acc_plot} and Table \ref{tab:bin_acc} show the percent difference between the predicted and true phases over the entire phase space (fractional error) vs. the number of observations, averaged over 6 runs. Our method surpasses the baselines in accuracy after 12 observations and achieves 60\% lower final error.

\textbf{Four phase problem}: Our model and Tian (SVC) were evaluated on a more complex phase diagram of four spaces inspired by the pressure-temperature phase diagram of materials (Figure \ref{fig:quad_pd} (\textbf{a})). Measurements are noiseless. Both models were initialized with 2 observations, so there are 2 phases hidden at the start so the model must discover the location of the two hidden phases. Figure \ref{fig:acc_plot} (\textbf{b}) shows the error vs. number of observations and final predictions are shown in Figure \ref{fig:quad_pd} (\textbf{b}) and (\textbf{c}).
Our model substantially outperforms the SVC after 22 steps, and perfectly predicts the phase diagram after 83 steps. SVC predictions are unaligned with observations, which was also observed in Ref~\cite{multiPhase}. Figure \ref{fig:ourPredictions} visualizes our model's predicted phase diagram, acquisition function, and error probability. Figure \ref{fig:gp_progress} in the Appendix shows the evolution of our model's predictions.


\section{Application on Phase Diagram of Magnetic Materials: Heisenberg Hamiltonian with Antisymmetric Interaction}
\begin{figure}
    \centering
    \includegraphics[width = 1.00\textwidth]{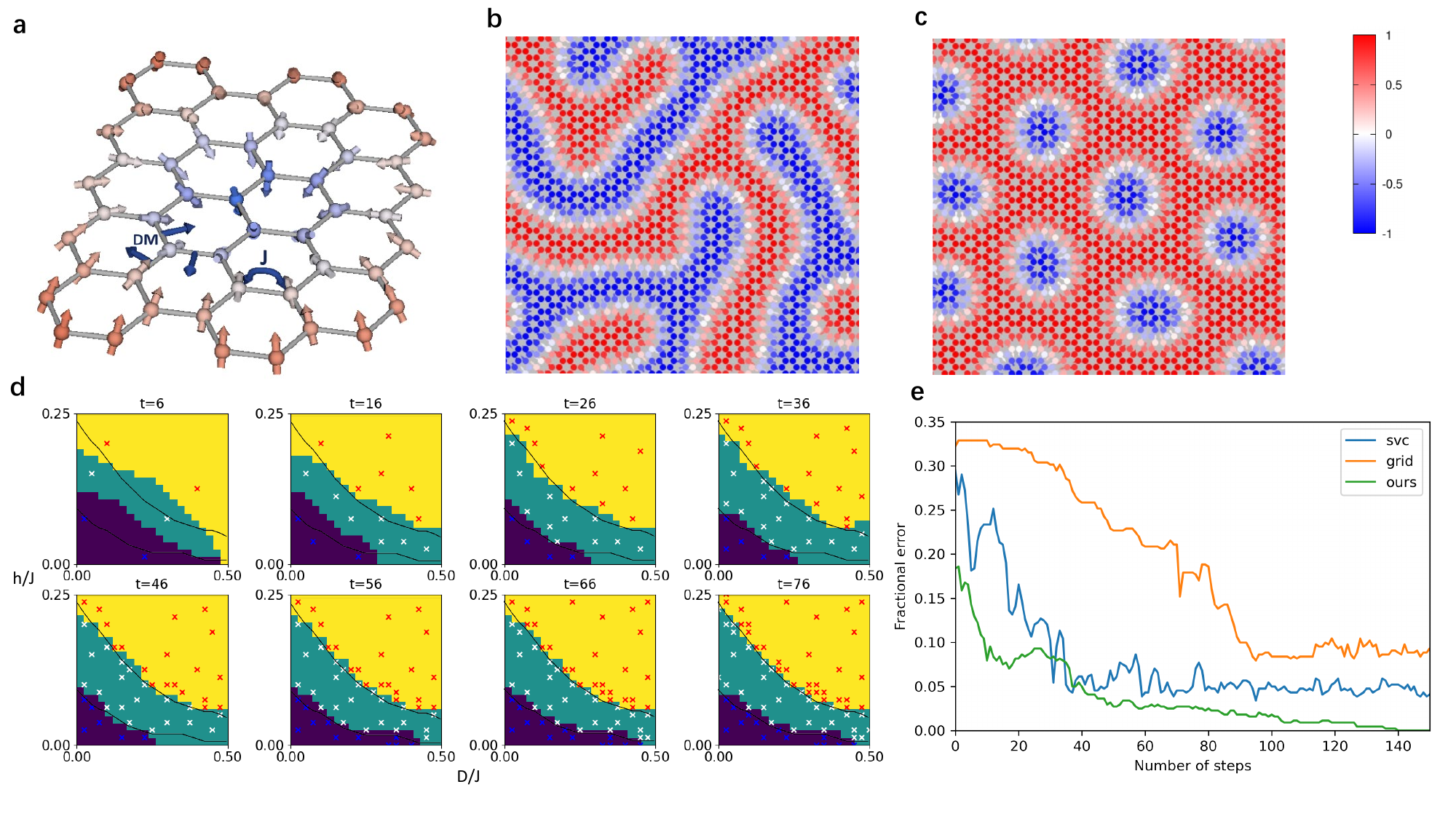}
    \caption{Phase diagram of Heisenberg model with antisymmetric interaction using our active learning approach. (\textbf{a}) A schematic of the model Hamiltonian and the microscopic spin configuration for skyrmion. $J$ and DM represent the spin exchange and DM vector for Dzyaloshinskii–Moriya interaction respectively. Single-ion anisotropy $D$ and magnetic field $h$ are not shown. (\textbf{b}) and (\textbf{c}) are real-space spin configurations of (\textbf{b}) spiral and (\textbf{c}) skyrmion phases. Color denotes $S_z$ component (colorbar shown) and arrows denote the spin direction in (\textbf{a}) and $(S_x, S_y)$ components in (\textbf{b}) and (\textbf{c}). Here (\textbf{b}) and (\textbf{c}) are obtained at $(D,h)/J = (0.056, 0)$ and $ (0.028, 0.7)$, respectively.  (\textbf{d}) Phase diagrams as functions of $D/J$ and $h/J$, for $d=0.5J$ and temperature $T=10^{-4}J$, obtained after $t$ observations. Crosses with colors represent points sampled: blue for the spiral phase, white for the skyrmion phase, and red for polarized phase. We use purple, green, and yellow to fill out inferred regions of the three different phases respectively. As a comparison, black lines show the ground-truth phase boundary. (\textbf{e}) Fractional error as a function of the number of observations (starting from $t=6$, with the same initial points) for SVC method (blue), grid sampling (orange), and our method (green).}
    \label{fig:honeycomb}
\end{figure}

A motivating application is the exploration of exotic phases in frustrated magnetic systems. 
One particular phase that has recently garnered significant attention is magnetic skyrmions. In the two-dimensional scenario, magnetic skyrmions manifest as real-space topological structures, formed by localized spin vectors that smoothly transition from their cores to their peripheries while maintaining fixed helicities. The real-space topology of a skyrmion is characterized by its skyrmion number $Q = \frac{1}{4\pi} \int \bm{m} \cdot (\frac{\partial \bm{m}}{\partial x} \times \frac{\partial \bm{m}}{\partial y}) dxdy$,  where $\bm{m}$ is the unit vector of local magnetization and the integral is taken over a two-dimensional space.\cite{topologicalnumber} To investigate the skyrmion phase, we focus on a Heisenberg model with antisymmetric interaction on the honeycomb lattice:
\begin{equation}
    \label{ham}
    H = -J \sum_{\langle i,j\rangle} \bm{S_i} \cdot \bm{S_j} -
    \sum_{\langle i,j\rangle}\bm{d_{i,j}}\cdot(\bm{S_i}\times \bm{S_j})-
    D\sum_{i} (S_i^z)^2 - h\sum_{i} S_i^z,
\end{equation}
where $\bm{S_i}$ represents the classical spin vector at site $i$ with $|\bm{S_i}|=1$ and $S_i^z$ is $z$-component. $J>0$ is the nearest neighbor (NN) ferromagnetic coupling, and $\bm{d_{i,j}}=d~(\bm{z}\times \bm{u_{i,j}}$) is the NN Dzyaloshinskii–Moriya (DM) interaction \cite{DMI1,DMI2}, where $\bm{z}$ and $\bm{u_{i,j}}$ are unit vectors, respectively perpendicular to the magnetic layer plane and pointing from site $i$ to site $j$, as shown in Fig. \ref{fig:honeycomb} (\textbf{a}). $D>0$ represents easy-axis single-ion anisotropy, and $h>0$ is the magnetic field. Besides the skyrmion phase, we have identified another two distinct magnetic phases in this system: the spiral and polarized phases, in which the spin configuration has magnetic wave vector $\bm{q}$ and $\bm{0}$. An example of the spiral phase and skyrmion phase is illustrated in Fig. \ref{fig:honeycomb} (\textbf{b}) (\textbf{c}). 

In order to determine the magnetic state configuration, we perform atomistic spin dynamics simulation by VAMPIRE~\cite{atomSpinSimulation} under fixed $J=1$ meV, $d/J = 0.5$ and temperature lowered to $T/J = 10^{-4}$. The system size is chosen to be $L_x=L_y=50a$ with periodic boundary conditions, where $a$ is the lattice constant. In constructing the phase diagram for $D/J\in[0,0.5]$ and $h/J\in[0,0.25]$ with $21\times 21$ discrete possible points using our algorithm, we start with 6 random initial starting points, then increase the number of observations to $t = 16, 26, 36, 46, 56, 66, 76$. The resulting phase diagrams are illustrated in Fig. \ref{fig:honeycomb} (\textbf{d}). As a comparison, the black lines in each plot represent the true phase boundaries obtained by grid sampling. The fractional error shown in Fig. \ref{fig:honeycomb} (\textbf{e}) is calculated by comparison with the true diagram via $21\times21$ grid sampling. The fractional error decreases as the number of observations grows, becoming smaller than $5\%$ after $t=37$ and smaller than $2.5\%$ after $t=60$, as illustrated in Fig. \ref{fig:honeycomb} (\textbf{e}). After $t=141$, the fractional error becomes zero, suggesting that our active learning approach now gives a perfect prediction for the entire phase diagram on the $21\times21$ grid in the two-dimensional phase space. Compared with direct grid sampling and the SVC method, our algorithm is much more efficient. Grid sampling is implemented by sampling points in order (sequentially sampling each parameter in order), starting from a half resolution grid before increasing to the full-resolution grid. Details are given in Appendix \ref{Appendix:baselines}. 

\begin{figure}
	\centering
	\includegraphics[width = \textwidth]{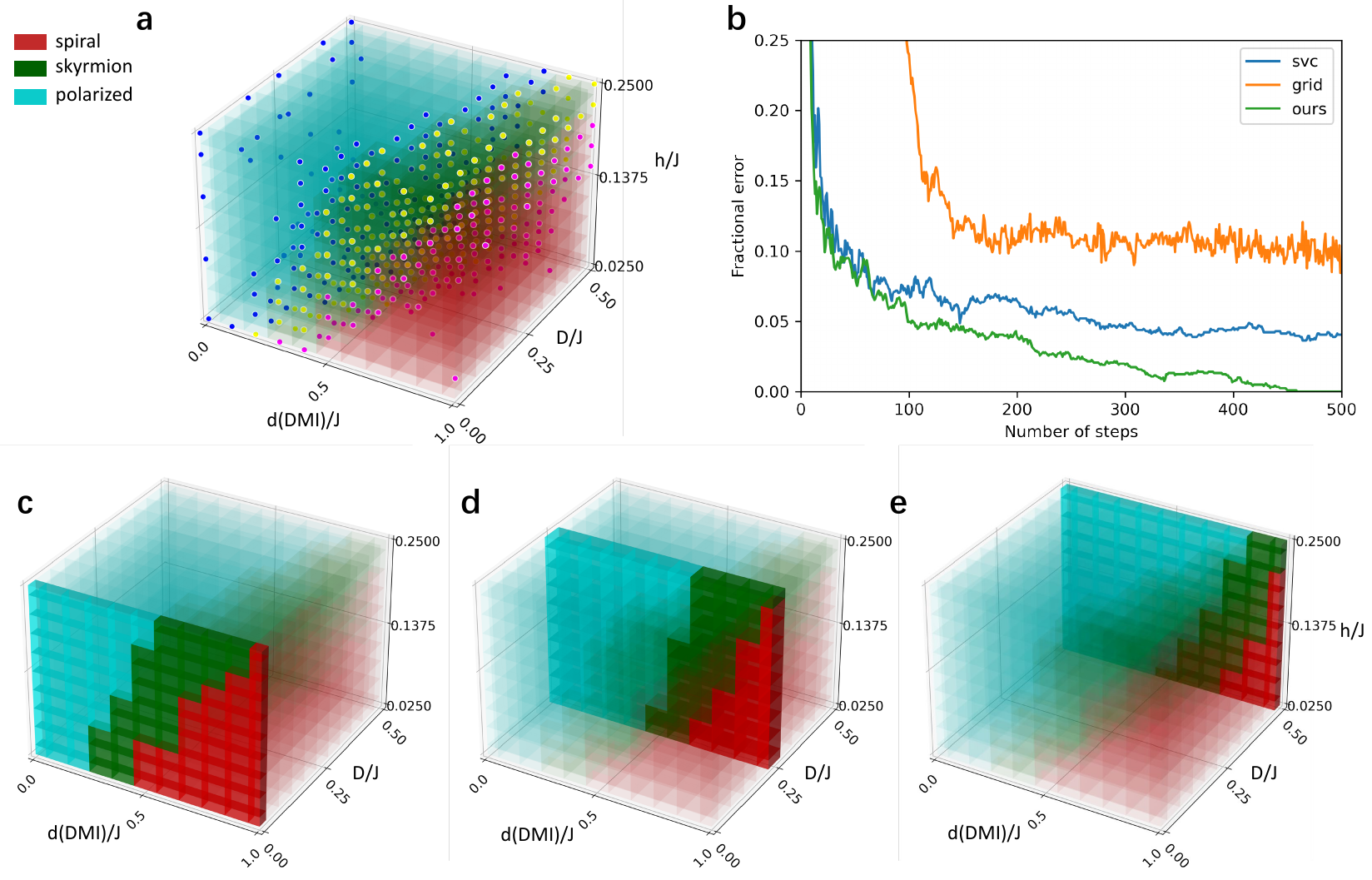}
	\caption{3D phase diagram of Heisenberg model with antisymmetric interaction using our active learning approach. (\textbf{a}) The regions of red, green, and cyan colors correspond to the spiral, skyrmion, and polarized phases, respectively. We also mark sampled observations with magenta, yellow, and blue little circles to represent the three phases, respectively. (\textbf{b}) Fractional error as a function of the number of observations (starting from $t=6$, with the same initial points) for SVC method (blue), grid sampling (orange), our method (green). (\textbf{c}) (\textbf{d}) (\textbf{e}) 3D phase diagrams that are highlighted at different values of $D$.}
	\label{fig:3D_pd_error}
\end{figure}

When more than two parameters are considered, the number of possible observations becomes enormous, making it very difficult to create a phase diagram using grid sampling. To further demonstrate the efficiency of our algorithm, we use it to generate a three-dimensional phase diagram for the Heisenberg model with varying antisymmetric interaction $d$ as mentioned above. Instead of setting $d/J = 0.5$, we now vary $d/J$ in the range of $[0,1]$, $D/J$ within $[0,0.5]$, and $h/J$ within $[0.025,0.25]$ with $11\times11\times10$ discrete possible points.

When $h=0$, the Hamiltonian exhibits at least two-fold degenerate ground states due to time-reversal symmetry. The classical spin simulation method we use can result in configurations with multiple domains, complicating the phase determination for polarized and spiral phases. Therefore, we exclude the case of $h=0$ in the 3D phase diagram analysis. The resulting phase diagrams are illustrated in Fig. \ref{fig:3D_pd_error} (\textbf{a}). The fractional error shown by Fig. \ref{fig:3D_pd_error} (\textbf{b}) green line is calculated by comparison with ground truth phase diagram via $11\times11\times10$ grid sampling. After around $t=99$, the fractional error becomes smaller than $5\%$, it converges to zero at around $t=440$.
The ratio of number of observations and number of total possible sample points is around $8\%$ to reach the fractional error smaller than $5\%$ for both 2D and 3D sampling. Compared to direct grid sampling and the SVC method, our algorithm demonstrates superior performance in both the rate of error reduction and the final error achieved.

To investigate the change of magnetic phases with varying antisymmetric interaction, single-ion anisotropy and external field, we snapshot a few 2D cuts of the 3D phase diagram in Fig. \ref{fig:3D_pd_error} (\textbf{a}) at easy-axis single-ion anisotropy $D/J=0, 0.25, 0.5$, as shown in Fig. \ref{fig:3D_pd_error} (\textbf{c}) (\textbf{d}) (\textbf{e}). Let us first focus on single-ion anisotropy $D=0$, a larger magnetic field $h$ favors the polarized phase, whereas stronger antisymmetric interaction $d$ causes the spins to form spiral structures. In the intermediate region, the competition between the magnetic field $h$ and antisymmetric interactions $d$ could stabilize the skyrmion phase. As the single-ion anisotropy $D$ increases, the phase boundaries shift to the right, corresponding to higher antisymmetric interaction $d$, and the polarized phase becomes more competitive. This means that the polarized phase can be stabilized at higher values of $d$ with increasing $D$. This behavior is expected because an easy-axis $D$ favors the alignment of spins in parallel directions (or antiparallel directions for other systems), requiring a larger antisymmetric interaction $d$ to disrupt the polarized phase. Concurrently, a larger antisymmetric interaction $d$ is needed to stabilize skyrmion phases as the single-ion anisotropy $D$ increases.

\section{Conclusion and Discussion}
Our proposed method is able to efficiently determine phase diagrams by identifying where to sample new observations, using a Bayesian approach to quantify the uncertainty. Demonstrated to be effective for synthetic experiments and for skyrmion phase diagram problem in both two-dimensional and three-dimensional phase space, this approach holds promise to significantly accelerate the identification of phase diagrams across a wide array of materials physics problems, even as the dimensionality of the parameters space increases. Our algorithm is shown to converge faster than grid sampling and other existing active learning methods. Additionally, our method can adapt to noisy measurements, making it particularly useful for experimental measurements where empirical errors are common. We provide our method as a package with source code and a graphical user interface for ease of use.

For future work, we would like to extend our approach to encompass more detailed prior information about the phase diagram. Users may have prior information about the shape, connectivity or other information about the phase diagram from before experiemnts are conducted, such as from prior experiments or theoretical work. In these cases, our Gaussain Process assumption that, a priori, phases are isotropically and uniformly Gaussian distributed would no longer hold true. Incorporating this information should further increase the efficiency of the active learning algorithm. 

To summarize, our active learning algorithm, if integrated with experimental or numerical methods for the automated sampling of phase diagrams, facilitates a higher level of automation with strategic and effective sampling in the phase space. We expect that this advancement will revolutionize scientists' and engineers' search for new materials and states of matter in high dimensional parameters space, allowing for more strategic exploration and unprecedented efficiency.

\section{Acknowledgments}
We thank Sathya Chitturi for the insightful discussion to the project. M. Zhu and Q. Zhao acknowledge support for this work from GSK, the Cambridge Centre for AI and Medicine, and the Engineering and Physical Sciences Research Council (Grant EP/V049968/1). C. J. Jia acknowledges the support from Department of Energy, Office of
Science, Basic Energy Sciences, Materials Sciences and
Engineering Division, under Contract No. DE-SC0022216, as well as the support from University of Florida Seed Grant. M. Mynatt acknowledges the support from the University Scholars Program at University of Florida. H. Pugzlys acknowledges the support from Christopher B. Schaffer undergraduate research scholarship.

\bibliographystyle{unsrtnat}
\bibliography{ref}

\section{APPENDIX}
\renewcommand{\thefigure}{A\arabic{figure}}
\setcounter{figure}{0}  

\begin{table}[h]
    \centering
    \begin{tabular}{cccccc}

\toprule
              &\multicolumn{5}{c}{Number of observations} \\
\textbf{Accuracy(\%)}      & 10 & 20 & 30 & 40 & 50\\
\midrule
Ours          &  0.071 & \textbf{0.038} & \textbf{0.028} & \textbf{0.021} & \textbf{0.018} \\
GPR                 &  0.139 & 0.055 & 0.051 & 0.052 & 0.033  \\
SVC                 &  \textbf{0.071} & 0.055 & 0.043 & 0.047 & 0.052 \\
\bottomrule
\end{tabular}
    \caption{Accuracy of predicted phase diagrams for binary classification after each algorithm after different numbers of observations. Results are averaged over 6 runs. The lowest error for each step is highlighted in \textbf{bold}. }
    \label{tab:bin_acc}
\end{table}

\subsection{Details of Baseline Algorithms} \label{Appendix:baselines}
In this section, we provide a detailed description of the three baseline algorithms.
\citet{efficientPhase} fit a Gaussian process (GP) is fit to binary observations $y = \{-1, +1\}$ and a heuristic acquisition function rates points, $\mathbf{x}$, to select, $A(\mathbf{x}) = \sqrt{var(y)} / (|E(y)| + \epsilon)$ where $var(y), E(y)$ and $\epsilon$ are the GP's variance, mean and a numerical stability factor. This acquisition function balances selecting points close to the currently predicted phase boundary where $E(y) \approx 0$ and regions of high uncertainty where $var(y)$ is large. The numerical stability factor prevents dividing by 0. Similar to our method, at every step, $A(\mathbf{x})$ is evaluated and the point with the highest value is selected as the next observation.

\citet{multiPhase} presents two methods. Method 2 is able to learn phase diagrams of more than two phases, however, this is not applicable to our setup since we cannot measure the location of phase transitions directly. Instead, we use a modified version of Method 1. A support vector classifier is fit to the observations of position and phase. Phase boundaries are predicted from classification boundaries. For active learning, we sample new points along the phase boundary that have the highest exponentially weighted average distance from existing samples, which acts as a proxy for regions of the highest uncertainty. This is because support vector classifiers cannot directly quantify prediction uncertainty, so we use distance from other observations as a proxy for uncertainty. We hand-tuned the exponential weighting factor for the best performance. Furthermore, since this model only places observations on the predicted boundary, it cannot discover disconnected phases that have not been observed. 

Grid sampling was implemented by sequentially sampling points starting from the origin, one line of points at a time. For example, in 2D, the column at $x=0$ is sampled, followed by $x=1$, for each $x$ coordinate. In 3D, the $z$ column $x=0, y=0$ is sampled, followed by $x=0, y=1$, for every $y$ column, after which the column $x=1, y=0$ is sampled until every point has been sampled. Since every point is sampled, this method generates a perfect phase diagram in the noiseless case. A downside of this grid sampling method is slow convergence since sampling starts from one direction. To mitigate this, instead of directly sampling from the full-resolution grid, the grid sampling algorithm first samples from a half-resolution grid and moves to the full-resolution grid after the half-resolution grid sampling is finished. In the 2D case (Figure \ref{fig:honeycomb}), the full grid is $21 \times 21$ so the starting grid was $11 \times 11$ and for the 3D case (Figure \ref{fig:3D_pd_error}), the full grid is $11 \times 11 \times 10$ so the starting grid was $6 \times 6 \times 4$. In both cases, the slowdown when switching from the low-resolution grid to the full-resolution grid is clear. Nearest-neighbour interpolation is used to make predictions for incomplete grids. 

\subsection{Implementation Details} \label{Appendix:details}
\textbf{Model details}: All of our Gaussian processes, $p(a_i|\mathbf{x})$, use the stationary Matern 1/2 kernel, $K(\textbf{x}, \textbf{x}') = K(r) = \kappa^2 exp(r/\rho)$, where $r$ is the Euclidian distance between $\textbf{x}$ and $\textbf{x'}$. The kernel parameters are $\kappa$, the magnitude of correlation between observations, and $\rho$, setting the lengthscale two observations are correlated. These are fitted using maximum likelihood estimation. We force the fitted kernel and noise parameters to remain positive and also clip them from becoming excessively large or small to stop numerical instabilities. The Matern 1/2 kernel generates predictions are continuous, in contrast with the Matern 3/2 or 5/2 kernels which also have (mean square) bounded derivatives. We chose the Matern 1/2 kernel since phase diagrams are discontinuous and do not vary smoothly. For further details about Gaussian Processes kernels, we recommend \citet{gpForML} Chapter 4. 

Our acquisition function assumes fantasy observations of phase that are distributed according to current model predictions (Equation \ref{eq:AcqFn}). We require $\hat{\epsilon}$, an estimate of the experimental error for fantasy observations. The true experimental error is unknown, so users can make a reasonable guess for the experimental error, e.g. based on previous observations. For our experiments, this is set to a constant $\epsilon=0.01$ for noiseless phase diagrams (4 phase diagram in Section \ref{sect:Synthetic} and Skyrmion phase diagram) or $\epsilon=0.10$ for noisy phase diagrams. Furthermore, for noiseless experiments, our algorithm is restricted to not search seen points again while noisy observations can be resampled multiple times.

\textbf{Numerical Optimisations}: To speed up our method, we make several numerical optimizations. Firstly, sampling probabilities $p(y|\mathbf{x}, \mathcal{T})$ (Equation \ref{eq:IntProb}) requires integrating Equation \ref{eq:Softmax} with respect to $n$ Gaussians, for each of the $n$ phases giving $n$ $n$-D integrals. Since probabilities sum to 1, this can be reduced to $n-1$ integrals. We efficiently and deterministically compute the integrals using the Gauss-Hermite quadrature instead of random sampling. This step is the most time-consuming, so it is performed in parallel using multi-processing. 

To find the next point $\mathbf{x}_{t+1} = \argmax_\mathbf{x} A(\mathbf{x})$, we sample points on a regular grid. Optimizations can be made to speed up the search. After the current probabilities of observing each phase are computed, we only need to sample $A(x)$ at points where there is significant uncertainty since very low probability observations are unlikely to impact the search. We set $A(\textbf{x})=0$ if the probability of observing any single phase is greater than 90\%. Also, since we are using a local kernel for our GPs, the effect of new observations is local so the distance $D(P_\mathcal{T}, P_{\mathcal{T} \cup \{\mathbf{x}, \mathbf{\hat{y}}(\mathbf{x})\}})$ is only sampled for local points within a set distance of $\mathbf{x}$ instead of the full area. This is set to be a grid of points with Euclidean distance within half the search range.

\begin{figure}
    \centering
    \includegraphics[width=\textwidth]{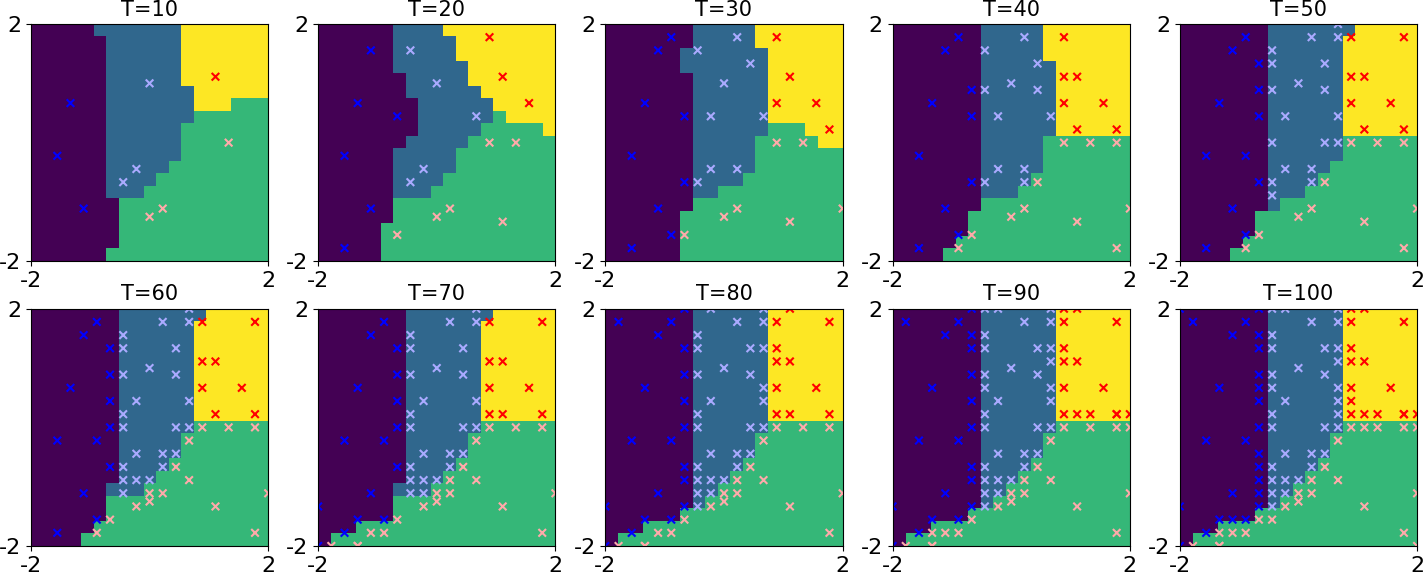}
    \caption{Observations and predictions of our model on the synthetic phase diagram of four phases after different numbers of observations. Observations are shown by crosses. The color indicates the observed phase.}
    \label{fig:gp_progress}
\end{figure}

\subsection{The criterion to determine phases of the honeycomb lattice}
The phases under different parameter sets can be easily identified by observing real-space spin textures (fig). We have also calculated some physical quantities to precisely identify the different phases. We employ the discrete version of topological charge $Q$ that identifies the existence of skyrmion phase. This version was first introduced in \cite{tpoChargeFirst} and used in \cite{tpoQused}. The calculation of $Q$ starts by triangulating the entire lattice and then counting the solid angels $\Omega_{\Delta}$ for each triangle $\Delta(\bm{S_1,S_2,S_3})$ determined by
\begin{equation}
    \label{solidangle}
    exp(i\frac{\Omega_{\Delta}}{2}) = \rho^{-1} [1+\bm{S_1\cdot S_2}+\bm{S_2\cdot S_3}+\bm{S_3\cdot S_1}+i\bm{S_1}\cdot(\bm{S_2}\times\bm{S_3})],
\end{equation}
where $\rho = [2(1+\bm{S_1\cdot S_2})(1+\bm{S_2\cdot S_3})(1+\bm{S_3\cdot S_1})]^{1/2}$, then $Q$ is given by summing over all of the triangles:
\begin{equation}
    \label{tpoQ}
    Q = \frac{1}{4\pi} \sum_{\Delta} \Omega_{\Delta}.
\end{equation}
Here the solid angle $\Omega_{\Delta}$ ranges from $-2\pi$ to $2\pi$ and the branch cut is the negative real axis when calculating $\Omega_{\Delta}$. We also calculate standard magnetization:
\begin{equation}
    \label{magetization}
    M = \frac{1}{N} \sum_i S_i^z,
\end{equation}
where $N$ is the number of sites in the entire lattice.

In Table. \ref{tab:criterion} , we show the criterion we use to assist the determination of phases.

\begin{table}[]
    \centering
    \begin{tabular}{|c|c|c|}
    \hline
       Phase & $Q$ & $M$ \\
    \hline
       Spiral & low or 0 & low \\
    \hline
       Skyrmion & high & high \\
    \hline
       Polarized & 0 & 1 \\
    \hline
    \end{tabular}
    \caption{Criterion used to assist the determination of three phases}
    \label{tab:criterion}
\end{table}

\subsection{Phase diagram obtained by grid sampling}
\begin{figure}
    \centering
    \includegraphics[width=0.5\textwidth]{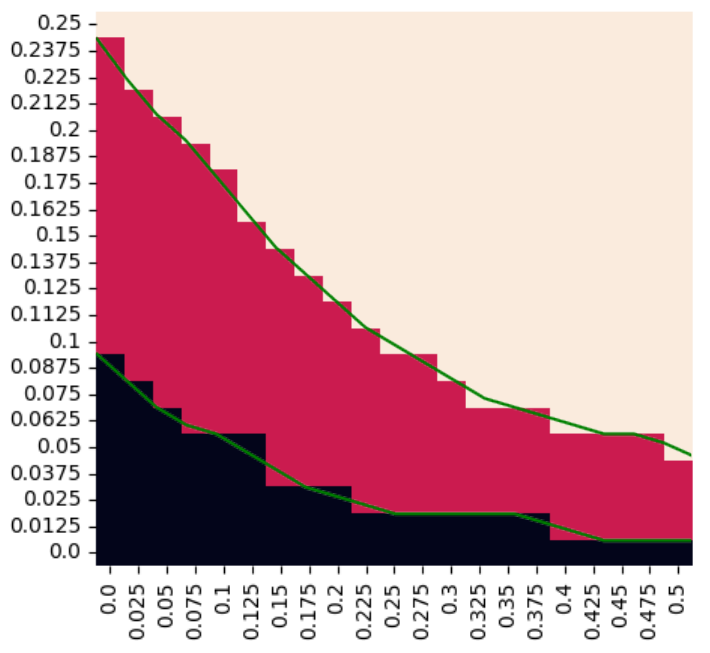}
    \caption{Phase diagrams of the honeycomb lattice as a functon of single-ion anisotropy $D/J$ (x-axis) and the magnetic field $h/J$ (y-axis) for fixed Dzyaloshinskii–Moriya interaction $d/J=1/2$ and temperature $T=10^{-4}J$ obtained by grid sampling. Black: spiral phase; red: skyrmion phase; pink: polarized phase. Green lines represent the smoothed phase boundaries.}
    \label{fig:PDgrid}
\end{figure}

\begin{figure}
    \centering
    \includegraphics[width=0.6\textwidth]{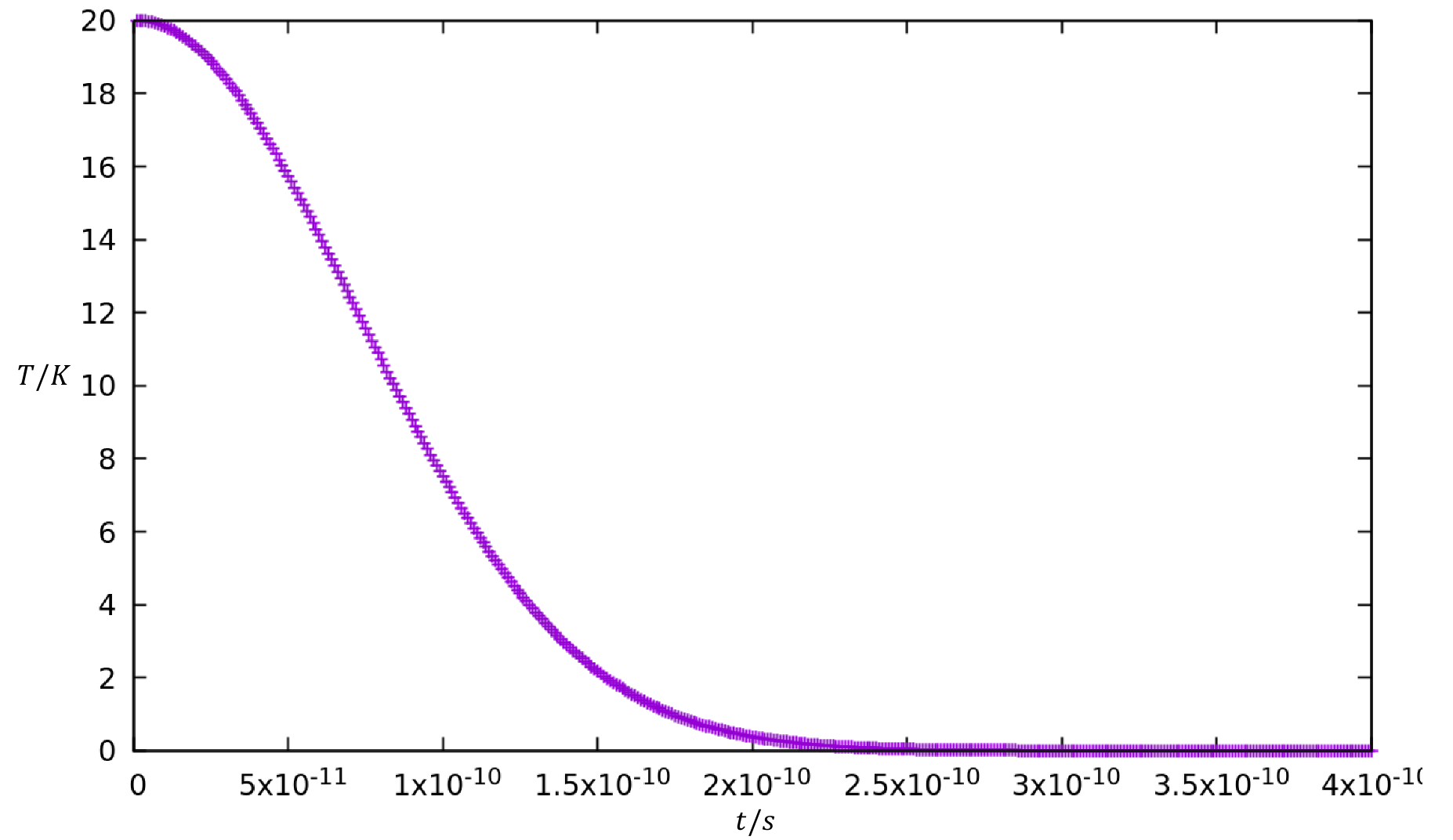}
    \caption{Temperature as a function of simulation time in our atomistic spin dynamics simulation.}
    \label{fig:vampireTvst}
\end{figure}

In Fig. \ref{fig:PDgrid}, we show phase diagram of the honeycomb lattice obtained by $21\times 21$ grid sampling and the phase boundaries (smoothed). To simulate the spin configuration of the honeycomb lattice, we simulate the atomistic spin dynamics using in Vampire \cite{atomSpinSimulation}. Starting from a relatively high temperature, we simulate the evolution of spins using the Landau–Lifshitz–Gilbert (LLG) equation \cite{LLG}, cooling to set temperature we set with custom timesteps. In this work, we cool the system from $T=20K$ to $T=10^{-4}J (0.00116K)$ (Fig. \ref{fig:vampireTvst}), at each temperature during cooling process, the system evolves $10^{4}$ steps with each step $10^{-16} s$, total $4\times 10^{6}$ steps in the whole process of a simulation. For the 3D case, we do $11\times 11\times 11$ grid sampling, and we cool the system from $T=30K$ to $T=10^{-4}J (0.00116K)$ when sampling.

\section{Additional information}

\subsection{Data Availability}
We provide code for our method at https://github.com/Maccy-Z/Active\_Learning\_Phase\_Diagram. This contains our algorithm and a graphical user interface written in Python for users, as well as code required to reproduce the synthetic experiments in Section \ref{sect:Synthetic}. 
\subsection{Competing Interests}
The authors declare that they have no competing interests.

\end{document}